\documentclass[11pt,tightenlines]{revtex4}
\usepackage{amsfonts}
\usepackage{amsmath}
\usepackage{txfonts}
\usepackage{graphicx}
\usepackage{amssymb}
\usepackage{color}
\usepackage{wrapfig,lipsum,booktabs}
\usepackage{hyperref}

\begin{document}
\title{Topological schemas of cognitive maps and spatial learning}
\author{A. Babichev$^{1}$, S. Cheng$^{2}$ and Yu. Dabaghian$^{1*}$}
\affiliation{$^1$Department of Neurology Pediatrics, Jan and Dan Duncan Neurological Research Institute, Baylor College of Medicine,
Houston, TX 77030 USA \\
Department of Computational and Applied Mathematics, 
Rice University, Houston, TX 77005 USA,\\
$^2$Mercator Research Group ``Structure of Memory'' and Department of Psychology, \\
Ruhr-University Bochum, Universitaetsstrasse 150, 44801, Bochum, Germany \\
E-mail: babichev@bcm.edu, sen.cheng@rub.de, dabaghia@bcm.edu$*$}

\date{\today}
\begin{abstract}
Spatial navigation in mammals is based on building a mental representation of their environment---a cognitive map. 
However, both the nature of this cognitive map and its underpinning in neural structures and activity remains vague. 
A key difficulty is that these maps are collective, emergent phenomena that cannot be reduced to a simple combination 
of inputs provided by individual neurons. 
In this paper we suggest computational frameworks for integrating the spiking signals of individual cells into a spatial 
map, which we call schemas. We provide examples of four schemas defined by different types of topological relations 
that may be neurophysiologically encoded in the brain and demonstrate that each schema provides its own large-scale 
characteristics of the environment---the schema integrals. Moreover, we find that, in all cases, these integrals are learned 
at a rate which is faster than the rate of complete training of neural networks. Thus, the proposed schema framework 
differentiates between the cognitive aspect of spatial learning and the physiological aspect at the neural network level.
\end{abstract}
\maketitle

\newpage

\section{Introduction}
\label{section:intro}
In the 1940's, Tolman proposed that animals build an internal representation---a cognitive map---of their environment 
and that this map allows the animal to perform space-dependent tasks such as navigating paths, finding shortcuts, and 
remembering the location of their nest or food source \cite{Tolman}. Three decades later, O'Keefe and Dostrovsky 
discovered pyramidal neurons in the hippocampus, named \textit{place cells}, that become active only in a particular 
region of the environment---their respective \textit{place fields} \cite{Best} (Figure~\ref{Figure1}A). The striking spatial 
selectivity of these place cells led O'Keefe and Nadel \cite{OKeefe1} to suggest that they form a neuronal basis of Tolman's 
cognitive map, thus providing this abstract concept with a concrete neurophysiological basis.
In the ensuing decades, it was realized that there are many brain regions involved in cognitive mapping of the environment 
\cite{Redish}, yet there is still no consensus on either the physiological mechanisms of this phenomenon or the theoretical 
principles that explain them \cite{McNaughton1}. Overall, it is believed that individual cells encode elements of the cognitive 
map, much like contributing pieces to a jigsaw puzzle. However, this analogy is not direct: the spiking activity of each separate 
neuron has no intrinsic spatial or geometrical properties---these properties appear only at the population level, emerging from 
the synchronous spiking activity of large neuronal ensembles \cite{Eichenbaum1,Pouget}. The mechanism of this phenomenon 
remains unknown, i.e., there exists a disconnect between the level of individual neurons from which the preponderance of 
neurophysiological data is acquired and the level of neuronal ensembles where the large-scale representations of space are 
believed to emerge \cite{Harnad}. 

In a recently proposed a model of spatial learning \cite{Dabaghian,Arai}, we attempted to bridge this gulf by combining recent 
experimental results pointing out the topological nature of the hippocampal map \cite{Alvernhe1,Alvernhe2,Poucet1,Shapiro,Wu,
Chen1,eLife} and methods of Algebraic Topology. This model allowed demonstrating that place cell activity can encode an 
accurate topological map of the environment and estimating the time needed to accumulate the required connectivity information. 
Further analyses of the model suggested to us that it is indicative of a more general theoretical framework that may lead to a 
systematic understanding of how spiking activity of neurons can be integrated to produce large-scale characteristics of space. 
In this paper, we outline the general principle and provide four specific models, which we call schemas, of integrating the activity 
of simulated neurons into a coherent representation of the explored environment. For each schema we find that a large-scale 
spatial map is produced within a short, biologically plausible period, which could be used to estimate the spatial learning time
in different environments.

\section{THEORETICAL FRAMEWORK}
\label{section:theory}
\subsection{The model} 

A schema model of a cognitive map contains the following three key components:
\begin{figure}
{\includegraphics[scale=0.86]{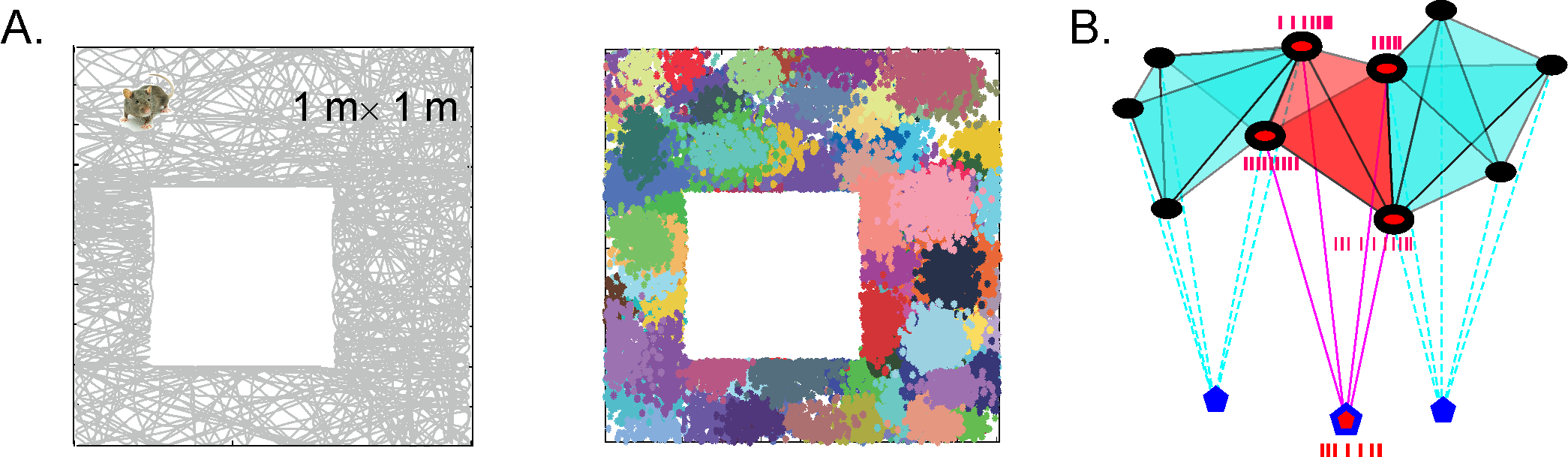}}
{\caption{\textbf{Physiological components of the schema}. \textbf{A}. The simulated trajectory in a $1\times 1$ m environment 
(left) and 200 randomly scattered place fields (clusters of colored dots) produced by place cells with a mean firing rate $f =12$ Hz 
and a mean place field size $s=20$ cm. \textbf{B}. Schematic representation of three overlapping assemblies of place cells (shown 
by black dots) that project synaptically onto their respective readout neurons (blue pentagons). The active place cells (black dots 
with red centers) of the ignited cell assembly (in the middle of the figure) produce spike trains that drive the spiking activity of the 
readout neuron (blue pentagon with the red center).}\label{Figure1}}
\end{figure}

\textbf{i. An abstract schema} $\mathcal{S}(R, P_S)$ represents the spatial information contained in the map at any given time. 
It consists of a set of formal regions $R = \{r_1, r_2, ..., r_N\}$ and a set of relationships, $P_S = \{\rho_1, \rho_2, ..., \rho_M\}$, 
that express how these regions combine. We presume that each region $r_i$ in the schema can be related to any other region 
$r_j$ through a chain of relationships with intermediate regions $\rho_{\alpha}(r_i,r_k)$, $\rho_{\beta} (r_k,r_l,r_m)$,..., 
$\rho_{\gamma }(r_n,r_j)$. A specific selection of the relationships included in $P_S$ determines the type of spatial information 
encoded in the schema and the global arrangement of the encoded regions, which is crucial both for the properties of the resulting 
map as well as for the information encoded in it.

\textbf{ii. The neural implementation}, $\mathcal{N}_S$, is a neural network that encodes the schema $\mathcal{S}$. For the sake 
of simplicity, we model $\mathcal{N}_S$ using a basic, two layer, feed-forward neural network inspired by cell assembly theory 
\cite{Buzsaki}, which consists of a layer of cells that represent regions of space and another layer of readout neurons that represent 
the relationships between these regions (Figure~\ref{Figure1}B). When a cell $c_i$ fires a spike, we say that the region $r_i$ 
is ``active''; otherwise it is ``latent'' \cite{Russell}. When a readout neuron fires a spike, we say that the corresponding relationship 
is ``instantiated.'' Thus, by construction, the relationships between regions are represented via temporal relationships between the 
spike trains and by the parameters of synaptic connections between the two layers in $\mathcal{N}_S$.

\textbf{iii. The spatial map and the representing space}. The goal of introducing schemas is to model the assemblage of the cognitive 
maps from the cells' spiking activity. However, in absence of a mechanism explaining how spatial representations emerge from the 
spike trains, this task remains undefined. Statements such as ``a given place cell's activity encodes a region'' or ``the coactivity of a 
set of place cells represents a spatial overlap between the encoded regions'' require an interpretation. In the analysis of 
electrophysiological data, this interpretation is acquired by mapping the neuronal activity into an auxiliary, external space $X$ which 
is selected according to the experimenters' best judgment. For example, constructing the place fields by ascribing Cartesian $x-y$ 
coordinates to the place cells' spikes and identifying the areas where the spikes cluster is one attempt to map the unobservable 
formal regions encoded in the cognitive map into observable regions of the spatial environment \cite{Barbieri}. In the following 
discussion, we will refer to this algorithm as to \textit{standard place field mapping}. Spaces that have been used to interpret the 
activity of place cells and other cells include Euclidean domains in one \cite{Frank1,Diba} and in three dimensions \cite{Hayman,Yartsev}; 
circles \cite{Taube}, tori \cite{Finkelstein}, spheres \cite{Chen}, and even Klein bottles \cite{Swindale}. To capture this aspect of 
cognitive map analysis, we define a \textit{spatial mapping} from the schema $\mathcal{S}$ to a representing space $X$,
\begin{equation}
f: \mathcal{S} \to X
\label{map}
\end{equation}
in which the formal regions of $\mathcal{S}$ are mapped into the ``concrete'' regions of $X$, $x_k = f(r_k)$. We will refer to $x_k$ 
as the $X$-\textit{representations of the formal region} $r_k$ and to the resulting layout of the representing regions in $X$ as the 
\textit{spatial} map of the schema, $M_X (\mathcal{S})$.

Although the representing regions are selected to reproduce the relationships between the formal regions as well as possible, the 
resulting map does not always capture the structure of the original schema: some relationships may be lost in the mapping or the 
mapping may produce relationships between the representing regions that are not encoded in $\mathcal{S}$. For example, the 
place field maps are believed to reflect an animal's cognitive map's structure but their faithfulness has not been established or even 
addressed in the neurophysiological literature. In the case when the set of relationships between the regions $x_k$ ($P_X$) matches 
the schematic relationships exactly, so that $P_X = P_S$, the mapping will be referred to as \textit{faithful}. The corresponding spatial 
map may then be viewed as a model of $\mathcal{S}$, i.e., the structure of $\mathcal{S}$ can be deduced from the layout of the 
representing regions. 

Thus, each specific schema model includes these three components---the abstract schema $\mathcal{S}$, its neuronal network 
implementation $\mathcal{N}_S$ and the spatial mapping (1) into a representing space $X$. For brevity, we will refer to this triad 
as to ``schema,'' when no ambiguities can arise. 

\subsection{Spatial learning} 
A key property of our approach, crucial for modeling spatial learning, is that schemas are dynamic objects. As an animal explores a 
novel environment, new regions become represented by the activity of place cells and new relationships are inferred from the spike 
trains' temporal patterns \cite{Dabaghian, Arai}. According to the standard approach of neural network theory, the process of learning a 
schema may be viewed as the process of training the readout neurons to represent the set $P_S$ by detecting repetitive patterns 
in the incoming spike trains. From this perspective, a schema is learned after its network is trained, i.e., after the readout neurons 
stop adopting their spiking responses to the patterns of the incoming spike trains.

On the other hand, from a cognitive perspective, the purpose of spatial learning is to acquire qualitative, large-scale characteristics 
of the environment, which enable spatial planning, spatial navigation and spatial reasoning, such as path connectivity, shortcuts and 
obstacles, geometric and topological properties, global symmetries and so forth. Such large-scale characteristics of the environment 
that are captured through the relationships of a given schema will be referred to as \textit{schema integrals}, $I_\mathcal{S}$. 
Below we demonstrate that the minimal time $T_{\min}$ required for the schema's integrals to emerge is typically shorter than the 
time, $T_\mathcal{N}$, required to train all readout neurons, i.e., large-scale information can be extracted from a partially trained 
network. Thus, the schema approach captures two complementary aspects of spatial learning: \textit{physiological learning}---the 
process of forming and training the cell assembly network and \textit{schematic} or \textit{cognitive learning}---the emergence of 
information about the global structure of space, expressed as the corresponding production of schema integrals.

\subsection{Topological Schemas}
What aspect of space is represented in the hippocampal map? The answer to this question depends on the information captured by the 
readout neurons in the hippocampal cell assemblies. Since correlating neuronal spiking with geometrical properties of the representing 
space sometimes produces useful interpretations of electrophysiological data, most authors assume that the spiking patterns of place 
cells encode geometric properties of space \cite{Barry,OKeefe2}. For example, it has been shown that combining the spiking activity of 
a relatively small number of place cells with the information about the sizes, shapes, and locations of their respective place fields allows 
a reconstruction of the animal's trajectory in a typical experimental enclosure on a moment-to-moment basis \cite{Guger,Brown1}. 

However, the read-out neurons have access only to the place cells' spikes, and not to their respective place fields. Obtaining the shape 
and size of any given place field, which is nothing but a cumulative spatial histogram of spikes used for illustrational purposes, requires 
accumulating a substantial number of spikes from the corresponding place cell. Yet the spike trains produced during the activity period 
of a given place cell are short---typically hundreds of milliseconds in duration---and highly variable, not only because of the animal's 
movements, but also because of the intrinsic stochasticity of neuronal spiking \cite{Fenton}. Thus, the spike trains of place cells contain 
little information about a place field, such as its shape, location and other computationally expensive parameters. Furthermore, recent 
experimental studies point out that these spike trains do not provide the geometric information on the synaptic integration timescale of 
seconds or fractions of a second \cite{eLife,Diba,Cheng1}.

Since the temporal pattern of place cell firing is the only information available to downstream neurons, a physiologically adequate class 
of schemas of the hippocampal map may be constructed based on capturing \textit{qualitative}, topological relationships between regions, 
e.g., overlap, adjacency, ordering and containment, from the temporal relationships between the spike trains \cite{Dabaghian1,Chen1}. 
The resulting maps will then produce a topological representation of space rather than a geometrical one \cite{Chen2,Stella}, in which the 
relative arrangement of the locations is more important than mapping the precise positions. Topological schemas have several advantages 
over the more precise geometric schemas, e.g., higher stability (e.g., faithfulness of a topological map is not destroyed under continuous 
deformations of the representing space) and lower computational cost, which may make them biologically more viable (see Discussion).

There remain many possibilities in which to read out qualitative information about the spike trains and thus there are many topological 
schemas. In this perspective, a particular readout mechanism, which responds to specific patterns of place cell coactivity, defines the 
type of spatial information encoded in the schema. The following discussion presents four different topological schemas based on different 
qualitative relations between regions and the rate at which these schemas are acquired.

\section{Results}
\label{section:results}

\subsection{Graph Schema $\mathcal{G}$}
\begin{figure}
{\includegraphics[scale=0.9]{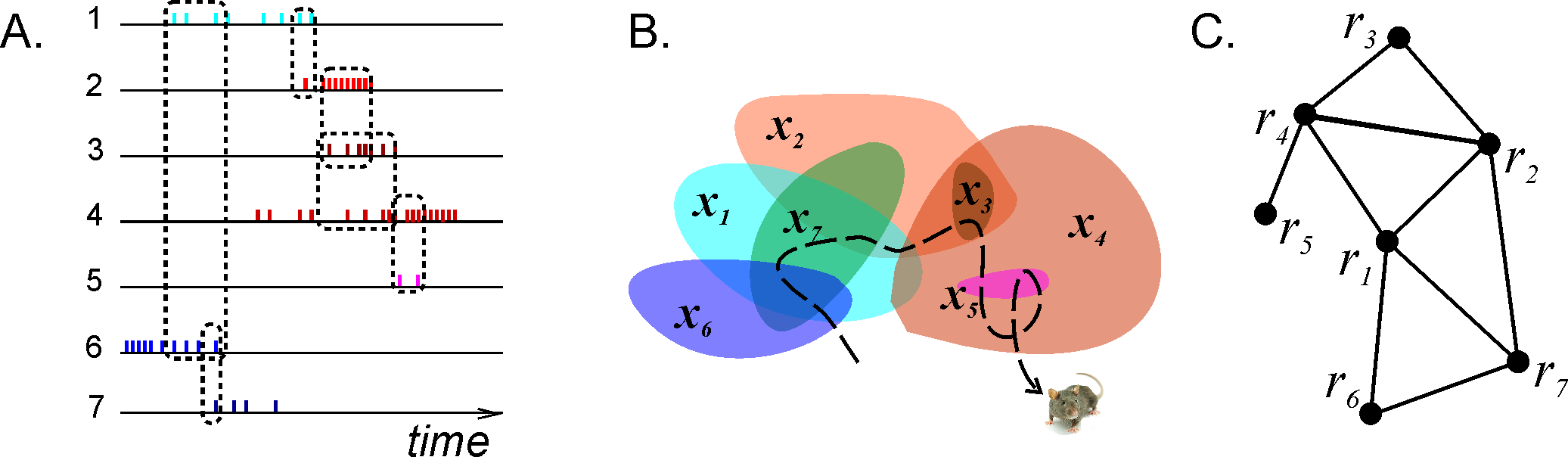}}
{\caption{\textbf{Graph schema}. \textbf{A}. A schematic illustration of the spike trains produced by seven place cells whose 
coactivity is indicated by the dashed rectangles connecting the spike trains. \textbf{B}. The corresponding seven place fields 
traversed by the animal's trajectory (dashed line). \textbf{C}. The corresponding graph schema, the seven vertexes of which 
correspond to seven formal regions encoded by the active place cells. The edges mark the relationships encoded in the schema, 
e.g., the connection $(r_4,r_5)$ is in the schema, but $(r_5,r_1)$ is not.}\label{Figure2}}
\end{figure}

The simplest topological schema is based on binary connections: its set of relationships consists of pairs of connected regions, 
$P_G =\{(r_i, r_j), (r_i, r_k), (r_m, r_n), ...\}$. Such schema can be viewed as a graph, $G$, whose vertices are linked if the 
corresponding regions are related according to $P_G$ (Figure~\ref{Figure2}). The corresponding neuronal implementation is 
produced by training the pair-coactivity detector readout neurons to respond to nearly simultaneous spiking of their respective 
pair of presynaptic cells \cite{Katz,Brette}. In other words, physiological learning of a graph schema $\mathcal{G}$ amounts to 
detecting pairs of cells that exhibit frequent coactivity \cite{cGraph}.

\begin{figure}
{\includegraphics[scale=0.9]{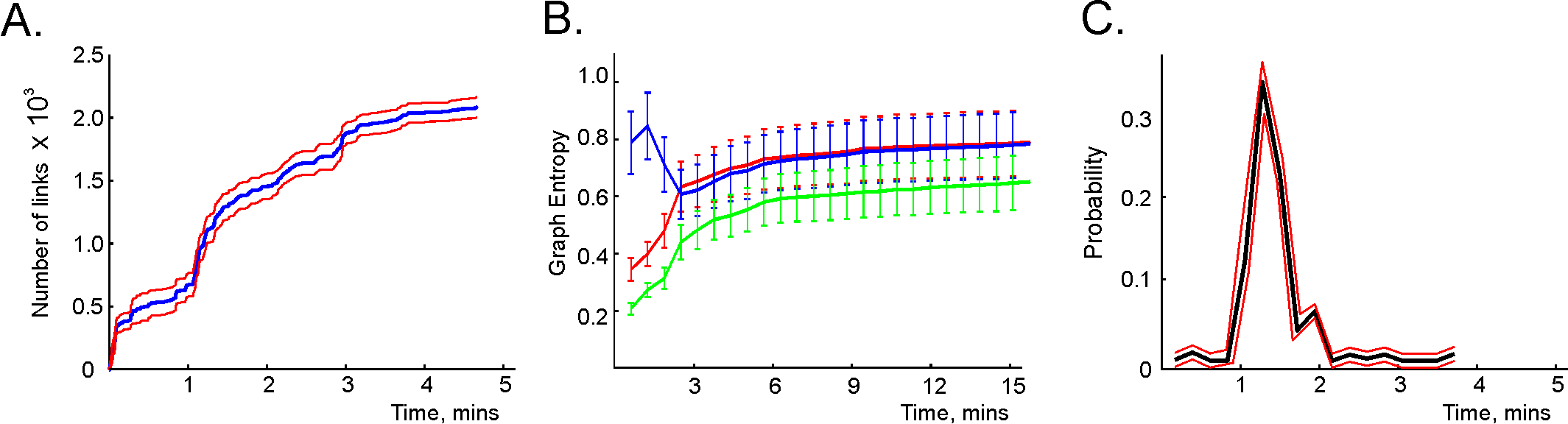}}
{\caption{\textbf{Spatial learning based on the graph schema}. \textbf{A}. The number of links in the graph schema as a 
function of time, computed for a simulated ensemble of 200 place cells with randomly scattered place fields (see Methods). 
The blue line represents the mean and the red lines show the error margins. \textbf{B}. Graph schema entropy (blue) and 
place field map entropy (red) as a function of time. The green line shows the mutual information between the map and the 
schema. Both entropies and the mutual information saturate at the time when the number of links saturates. \textbf{C}. 
The probability of establishing a maximal length connection in the graph schema stabilizes in about 2.2 minutes, when only 50\% 
of connections appeared.}\label{Figure3}}
\end{figure}

We modeled this process by simulating place cell spiking activity induced by a rat's movements across a place field map in a 
small environment (Figure~\ref{Figure1}B). To simplify the analyses, we assume that as soon as the coactivity occurs, the 
corresponding connection is immediately ``learned,'' i.e., incorporated into the schema. As a result, at every moment of time $t$, 
the connectivity matrix of the graph is defined by the coactivity observed prior to that moment. Thus, $C_{ij} = 1$ if cells $c_i$ 
and $c_j$ cofired before t and $C_{ij} = 0$ otherwise. Figure~\ref{Figure3}A shows that the number of links in the graph, 
which is the number of recruited pair-coincidence detectors, grows as the schema is learned and saturates at ca. 
$T_\mathcal{N}  = 5$ mins, i.e., after this time new incoming spike trains do not produce new connections in $\mathcal{N}_G$. 

The saturation of the schema could be a trivial result if the graph becomes fully connected or remains mostly empty. A simple 
characteristic capturing the efficiency of $\mathcal{G}$, which generalizes to other schemas in a natural way is its entropy 
\cite{Dehmer,Mowshowitz}. This is the specific entropy of the readout neurons,
$$H_G = - p_c  \log_2(p_c) - p_d \log_2(p_d),$$
where $p_c$ and $p_d = 1- p_c$ are the fractions of the connected and disconnected vertex pairs in the graph. For a fully discrete 
($p_c = 0$) or a fully connected graph ($p_c = 1$) the entropy vanishes and maximal entropy $H_G = 1$ is achieved for $p_c = 1/2$ 
(in which case the absence of a link is as informative as its presence). Figure~\ref{Figure3}B demonstrates that for the place cell 
ensemble used in our simulations (see Methods), the entropy of the graph schema asymptotically approaches a maximal 
value of about $H_G \approx 0.8$ in about five minutes, a value implying that the schema network $\mathcal{N}_G$ is neither 
underloaded nor oversaturated. 

To quantify the correspondence between the schema $\mathcal{G}$ and its place field map $M_X(\mathcal{G})$, we calculated the 
entropy $H_X$, of the occurrences of place field pairwise overlaps across time and compared $H_X$ to $H_G$. Figure~\ref{Figure3}B 
demonstrates that both entropies remain close throughout the entire learning period, indicating that the complexity of the place field 
layout remains similar to the complexity of the encoded relationships at all times. In addition to this correspondence we also computed 
the mutual information (MI, see Methods) between the place field overlap and place cell coactivities, which also grows with the rat's 
navigational experience (Figure~\ref{Figure3}B). Thus, we have convergent lines of evidence indicating efficient spatial learning 
captured by the graph schema.

As a cognitive map model, the graph schema $\mathcal{G}$ provides a stratum for implementing graph-theoretical navigation 
algorithms, that is, for establishing paths connecting spatial locations \cite{Trullier,Chrastil}. Its integrals $I_{\mathcal{G}}$ are 
the global characteristics of the region-to-region connectivity graph, e.g., its partitioning, the colorability of its vertexes and edges 
\cite{Berge}, its planarity, and the existence of a path between two given vertexes. As an example of such large-scale characteristics 
we identified the shortest paths connecting pairs of the most distant vertexes and computed the time required to establish these 
connections. The results shown in Figure~\ref{Figure3}C demonstrate that the animal establishes connections between the 
most distant locations in the graph in about $T_{\min} = 2.2$ minutes, a time when only about $50 \% $ of the readout neurons 
are trained. Similarly, emergence of the information required to establish existence of a circuit of the graph which traverses each 
edge exactly once, called an Eulerian path, takes about $2.2$ minutes, while the correct number of cliques in $\mathcal{G}$, which 
are sets of pairwise connected vertices, can be deduced within two minutes. These observations suggest that the emergence of 
schema integrals before the network is trained may be a general phenomenon.

\subsection{Higher-order Overlap (Simplicial) Schema $\mathcal{T}$}
\begin{figure}
{\includegraphics[scale=0.9]{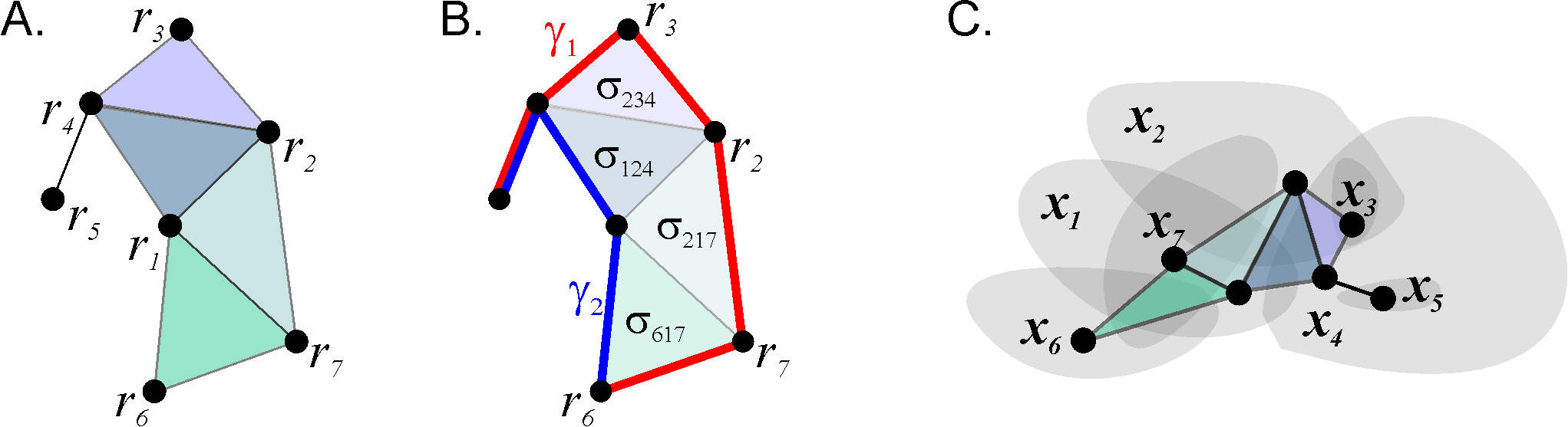}}
{\caption{\textbf{Simplicial schema}. A schematic representation of the connections between the vertexes associated with the seven 
place cells and their spatial map shown on Figure~\ref{Figure1}B. \textbf{B}. The existence of the two-dimensional simplexes, 
corresponding to higher order coactivity relationships, permits the links to be deformed. This is illustrated for paths $\gamma_1$ and 
$\gamma_2$: the transformation can be visualized by slipping one path across the two-dimensional facets, thus demonstrating the 
topological equivalence between paths $\gamma_1$ and $\gamma_2$. \textbf{C}. The nerve of the map shown on Figure~\ref{Figure1}B 
matches the simplicial schema.}\label{Figure4}}
\end{figure}

A topological schema may be based on representing not only binary, but also ternary, quaternary and other higher-order connectivity 
relations between spatial domains. For example, a schema may represent the overlaps between regions, including triple, quadruple, 
etc., overlaps. The key property of the overlap relation is that if $k$ regions, $r_1$, $r_2$, ..., $r_k$, have a common intersection, then 
so does any subcollection of them. The simplest mathematical object that is closed under the operation of taking non-empty subsets is 
an abstract simplex, which can be viewed as a list of $k$ elements \cite{Aleksandrov}. Hence, a $(k+1)$-order overlap relationship 
$\rho(r_0,r_1,...,r_k)$ may be represented by a $k$-dimensional simplex $\sigma = [r_1, r_2, ..., r_k]$. 
A set of overlap relationships therefore forms an abstract simplicial complex, $\mathcal{T}$, and we will hence refer to a higher order 
overlap schema as to simplicial schema. 

Under the standard mapping of the place cell spiking activity into the environment, the simplicial schemas' relationships, $P_{\mathcal{T}}$, 
represent the overlaps between the place fields. For example, the place field map shown in Figure~\ref{Figure2}B can be faithfully encoded 
by a simplicial schema with four $3^d$ order relationships $P_3 =\{(r_6,r_1,r_7), (r_7,r_1,r_2), (r_1,r_2,r_4), (r_2,r_3,r_4)\}$ and an 
additional binary relation $(r_4,r_5)$, as shown in Figure~\ref{Figure4}. The neuronal marker of these overlaps is the spiking coactivity: 
if the animal enters a location where several place fields overlap, their respective place cells produce (with a certain probability) temporally 
overlapping spike trains. Hence the neural network implementation of a simplicial schema, $\mathcal{N}_{\mathcal{T}}$, should be built 
to detect the coactivity events, using coincidence detector readout neurons (which, in fact, corresponds to the current view on the hippocampal 
cell assembly network organization \cite{Buzsaki,Harris1,Harris2, Babichev}). 

\begin{figure}
{\includegraphics[scale=0.8]{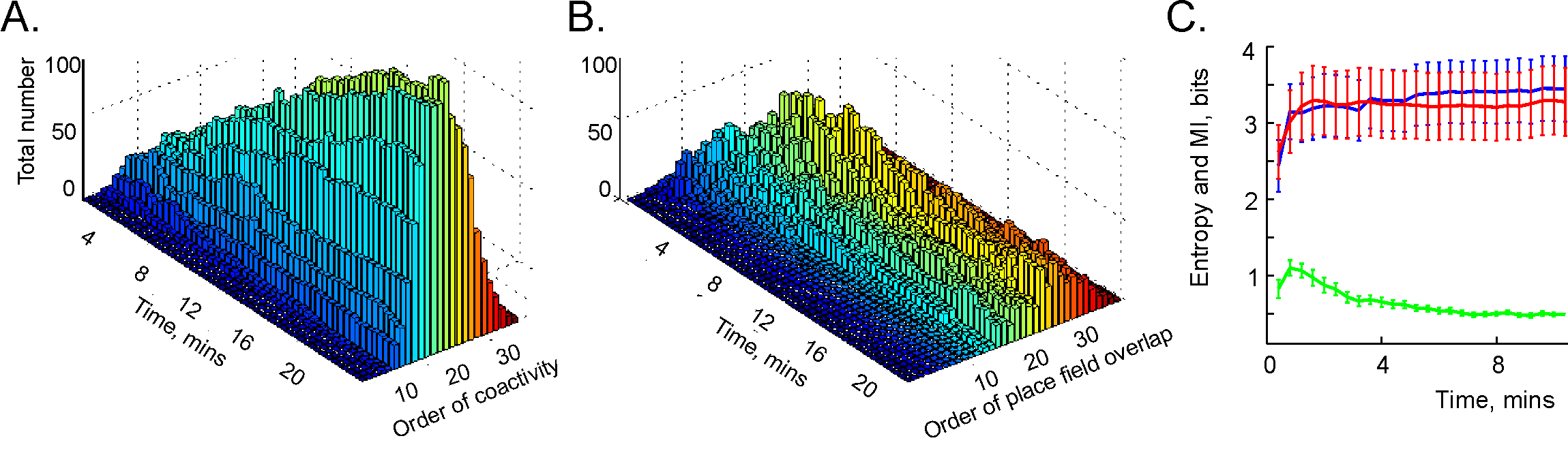}}
{\caption{\textbf{Learning the simplicial schema}. \textbf{A}. The development of the population of readout neurons as a function of time. 
The typical order of co-activity is about 20, the highest order is 33. The population of the high order readout neurons (order above 11) 
increases through the entire duration of the experiment (over 20 minutes). Other populations reach a stable plateau (e.g., orders 9 and 10) 
or reach a maximum and then drop (e.g., order under 8). The decrease of the number of the low order relationships after an initial increase 
indicates the elimination of redundant information. \textbf{B}. The development of the overlap relationships between the standard representing 
regions as a function of time. The typical overlap order is about 20, the highest order is 35. As in the case with the readout neurons, the 
number of high-order overlaps (order above 17) saturates on the rise, unlike the lower order overlaps. \textbf{C}. The entropy of the dimensions 
of registered simplexes (red) is similar to the entropy of the orders of the overlaps of the concrete regions $x_k$ (blue). The mutual information 
between these two variables is computed along the trajectory (green).}\label{Figure5}}
\end{figure}

Physiological learning of a simplicial schema hence amounts to training the readout neurons to detect place cell coactivities. Our learning 
algorithm (see Methods) ensures that, at every stage of learning, only the highest order relationships are kept while the redundant lower-order 
relationships are eliminated. For example, pairwise connections between three neurons become redundant after a triple coactivity between 
them is detected, at which point the three pair-detector readout neurons can be replaced with a single triple-coincidence detector. Numerical 
simulations demonstrate that, as the rat explores the environment, the more probable, lower-order coactivity events are captured first and 
the less probable higher order coactivities accumulate more slowly (Figure~\ref{Figure5}A). Moreover, although rapid changes of the readout 
neurons' order stops after five or six minutes, slow regroupings continue during the entire navigation period, $T = 25$ minutes. Thus, unlike 
the pairwise overlaps in $\mathcal{G}$ that can be instantly identified, the orders of the readout neurons cannot be deduced from a single 
coactivity event. In this sense, the orders of the readout neurons are integral characteristics of place cells' spiking activity, and therefore may 
be viewed as integrals of the simplicial schema.

There exists an additional important set of $\mathcal{T}$-integrals, which capture the topology of the representing space. This property 
of the simplicial schemas can be illustrated using the \v{C}ech theorem, which states that the pattern of overlaps between regions $U_1$, 
$U_2$, ..., $U_n$, covering a topological space $X = \cup_i U_i$, encodes homological invariants of $X$, provided that every intersection 
$U_i \cap U_j \cap ... \cap U_k$ is contractible \cite{Hatcher, Dubrovin}. The proof is based on building the ``nerve'' of the covering---a 
simplicial complex, the $d$-dimensional simplexes of which correspond to the $(d +1)$-fold overlaps between covering regions, and showing 
that it is topologically equivalent to $X$ (Figure~\ref{Figure4}C). This theorem implies that the spatial map of a sufficiently rich simplicial 
schema may encode the topology of the space navigated by the rat, and suggests that if this map is faithful, i.e., if the nerve of the spatial 
map matches the schema's relationship set $P_{\mathcal{T}}$ exactly, then the schema also captures the large-scale topological representation 
of the space. 

To study the correspondence between the simplicial schema and its map, we compared the schema's entropy $H_{\mathcal{T}}$, defined 
by the probabilities for a readout neuron to become a $k^{th}$-order co-activity detector, to the entropy $H_X$ of the place field map 
$M_X(\mathcal{T})$, defined via the probabilities of producing a $k^{th}$-order overlap between the place fields (Figure~\ref{Figure5}B). 
As shown on Figure~\ref{Figure5}C, both entropies closely follow one another: they both grow initially and reach similar asymptotes in 
approximately four minutes. However, the mutual information between these two series of events decreases with time. The reason for this 
effect lies in the idealized nature of the representing regions $x_k$, built as convex hulls of the spike clusters in the two-dimensional environment 
(for other place field construction algorithms see \cite{Maurer,Muller1}). The $x_k$'s crisp boundaries produce high-order overlaps, which 
are not captured by the place cell coactivity and hence by the schema---compare the orders of the readout neurons on Figure~\ref{Figure5}A 
with the orders of overlap between the corresponding representing regions $x_k$ in Figure~\ref{Figure5}B.
\begin{figure}
{\includegraphics[scale=0.8]{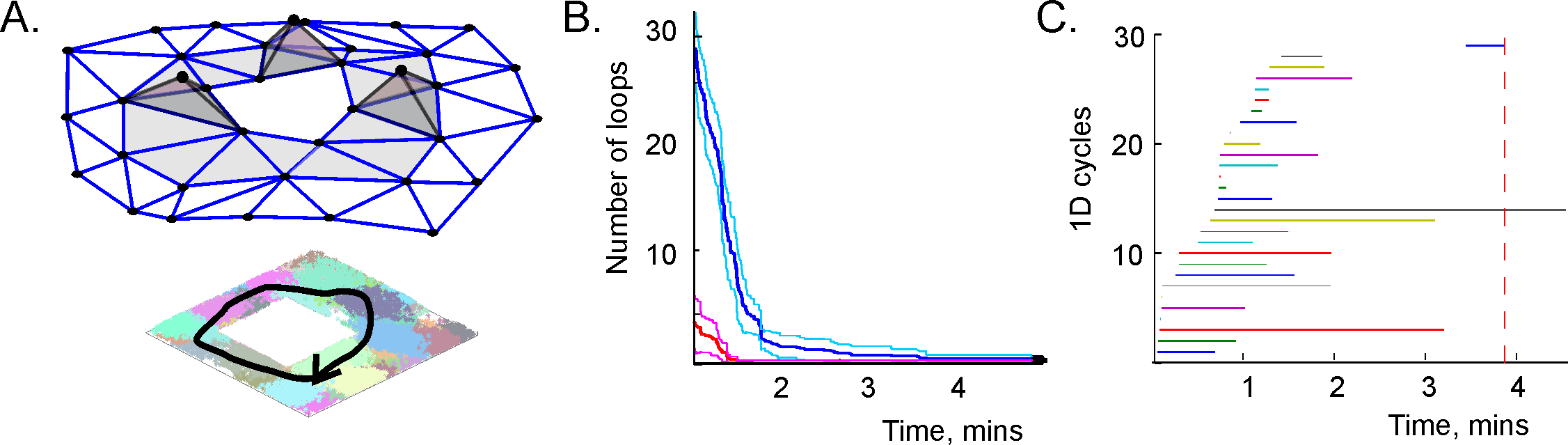}}
{\caption{\textbf{Topological loops in simplicial schema}. A sequence of the place cell combinations ignited along a path $\gamma$ (black line) 
corresponds to a sequence of simplexes---a simplicial path $\Gamma$ that represents $\gamma$ in $\mathcal{T}$. \textbf{B}. The dynamics of 
the total number of zero-dimensional loops (red) and one-dimensional loops (blue). Unlike the growing number of links in the graph schema 
(Figure~\ref{Figure2}A), the number of topological loops decreases with time. Eventually, a single loop survives. The margins of error are 
shown above and below each graph by a pair of pink and light-blue lines, respectively. \textbf{C}. The barcodes---timelines of the one-dimensional 
loops in the simplicial complex. The topological noise vanishes after ca. 4 minutes, which is the schematic estimate of the cognitive learning time.}
\label{Figure6}}
\end{figure}

This result can be viewed from several perspectives. First, it illustrates that the scope of reliable information that can be drawn from the spatial 
map is limited: only sufficiently robust, qualitative aspects of the place field map, such as low dimensional overlaps, can be trusted. Second, 
the regions $r_i$ that were originally introduced as ``formal,'' that is, devoid of intrinsic properties, should fundamentally be viewed as ``fuzzy'' 
and not as Euclidean domains with crisp boundaries \cite{Liu}. 

Direct computations show that the coactivity complexes do, in fact, capture the topology of the representing space, provided that the place cells' 
spiking parameters fall into the biological domain \cite{Dabaghian, Arai, Babichev}, and hence that simplicial schemas provide a framework for 
representing topological information. For example, cell assemblies ignited along the physical paths traversed by the animal correspond to 
sequences of coactivity simplexes---the simplicial paths that represent the physical paths in $\mathcal{T} $ (Figure~\ref{Figure6}A). The 
structure of the simplicial paths allows establishing topological (in)equivalences between navigational paths, e.g., topologically equivalent 
simplicial paths represent physical paths that can be deformed into one another, a non-contractible simplicial path corresponds to a class 
of the physical paths that enclose inaccessible or yet unexplored parts of the environment. As a result, the simplicial schema produces a 
qualitative description of navigational routes: while the total number of paths grows exponentially, the number of topologically distinct loops, 
which represent topologically distinct paths is small (Figure~\ref{Figure6}B). 

However, this information does not emerge immediately: as the animal begins to navigate a new environment, most topological loops reflect 
transient connections. As the spiking information accumulates, these ``spurious'' loops disappear and only the loops that correspond to the 
physical signatures of the environment persist (Figure~\ref{Figure6}C). 
With methods drawn from persistent homology theory \cite{Zomorodian,Ghrist} one can determine the minimal period $T_{\min}$ required 
for removing the spurious loops, which provides a theoretical estimate of the time required to learn the environment \cite{Dabaghian, Arai}. Figure~\ref{Figure6}C demonstrates that in our test map, after $T_{\min} = 4$ minutes most topological loops have vanished and only the 
loops that correspond to the physical holes in the environment survive. Thus, as in the case of the graph schema, the topological connectivity 
of the cognitive map is captured by the simplicial schema before the underlying neuronal network is fully trained, $T_{\min} < T_{\mathcal{N}}$. 

\subsection{Mereological Schema $\mathcal{M}$}
\begin{figure}
{\includegraphics[scale=0.9]{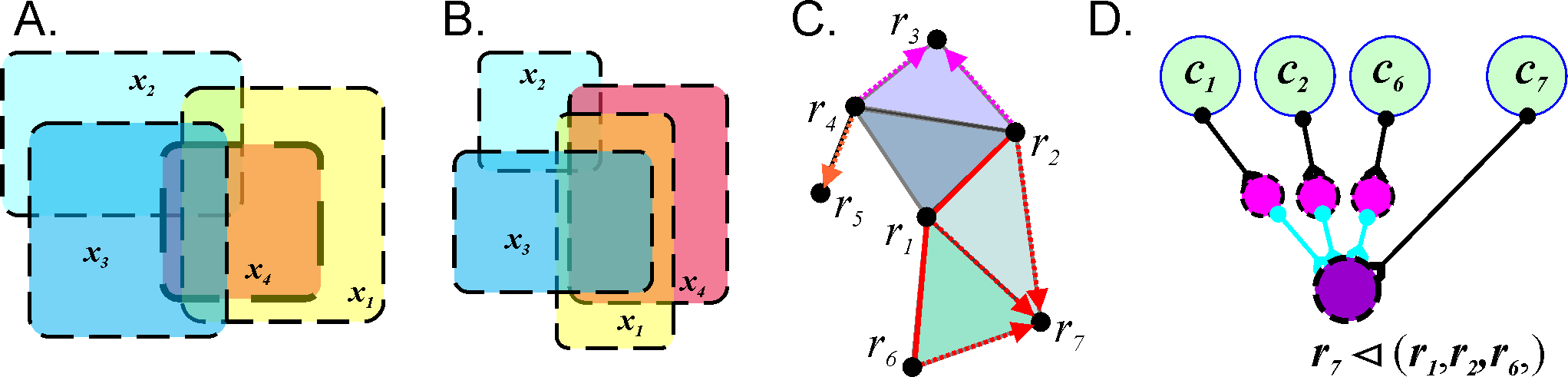}}
{\caption{\textbf{Mereological schema}. In \textbf{A} and \textbf{B}, the overlap pattern does not capture the cover relationship. Four 
regions, $x_1$, $x_2$, $x_3$ and $x_4$ form a quadruple overlap in both cases. However, in map shown in \textbf{A}, the region $x_4$ is 
contained in the union of $x_1$, $x_2$ and $x_3$. In the map shown in \textbf{B}, $x_4$ is not covered. \textbf{C}. The cover and the overlap 
relationships in a mereological schema corresponding to the map of Figure~\ref{Figure1}A. The covering regions are connected by 
red links (e.g., $r_6$, $r_1$, and $r_2$) and the red arrows point to the covered region (e.g., $r_7$). \textbf{D}. A neuronal implementation 
of the covering relationship includes three inhibitory neurons (magenta) that provide inhibitory input into the readout neuron (purple). If each 
inhibitory input of an active interneuron exceeds the excitatory input of the driving cell $c_7$, the readout neuron can spike only if the activity of 
the cell $c_7$ is not accompanied by the inputs from any of the cells $c_1$, $c_2$ or $c_6$. As a result, the readout neuron will remain silent 
as long as the activity of the cells $c_1$, $c_2$ and $c_6$ temporally covers the activity of cell $c_7$.}\label{Figure7}}
\end{figure}

Although a sufficiently rich simplicial schema can capture the topological invariants of the representing space $X$ as its integrals, it does not 
capture all the qualitative topological aspects of the connectivity between regions. For example, the identical simplicial schema (represented
by a tetrahedron) can faithfully represent the overlap relationships in the two maps shown in Figure~\ref{Figure7}, because both maps contain 
the same set of regions $R = \{r_1, r_2, r_3, r_4\}$ and one fourth-order overlap relation $P_R = \{(r_1, r_2, r_3, r_4)\}$, as well as all 
their consequent ternary and binary sub-relations. However, these maps are topologically different, since they cannot be transformed from 
one into another by a continuous deformation of the plane $R^2$. The obstruction to such deformation is that the region $x_4$ on 
Figure~\ref{Figure7}A is contained in the union of the regions $x_1$, $x_2$ and $x_3$, i.e., $x_4 \subset (x_1 \cup x_2 \cup x_3)$, and no 
containment relationships exist between any combinations of the regions on Figure~\ref{Figure7}B. Neither a graph schema $\mathcal{G}$ 
nor a simplicial schema $\mathcal{T}$ can capture this difference; what is required is the additional covering relation, $(x_1, x_2, ..., x_l) 
\blacktriangleleft (y_1, y_2,..., y_k)$, ($x'$s are covered by $y'$s), in terms of which the map on Figure~\ref{Figure7}A is described by the 
relationship $r_4 \blacktriangleleft (r_1, r_2, r_3)$, whereas the regions shown on Figure~\ref{Figure7}B produce no containment relation. 
The cover relation produces a new---mereological---schema $\mathcal{M}$, in which the information is encoded in terms of topological 
containment (Figure~\ref{Figure7}C). The intuition behind neuronal implementation of the formal cover relation is the following. If the activity 
of one ensemble of place cells, $U = \{c_1, c_2, ..., c_k\}$, outlasts, or covers in time, the activity of cells in another ensemble 
$V = \{d_1, d_2, ..., d_l\}$, then the region $X_U$, representing the $U$-ensemble, contains the region $X_V$ representing the $V$-ensemble:
$$X_V \subset X_U  \,\,{\rm  if }\,\,  V \blacktriangleleft U.$$
From this perspective, the set of covering cells provides contextual information about the covered cells, i.e., the cover relation combines the 
basic formal regions into more complex, composite regions.

The cover relationship can be implemented, e.g., by a combination of the excitatory and inhibitory neurons shown on Figure~\ref{Figure7}D. 
In such a cell assembly, the readout neuron signals a violation of the cover relationship, i.e., the latter is represented by an absence of the 
readout neurons' spiking activity up to the moment $t$. Hence, in contrast with simplicial schemas, where readout neurons learn to detect ever 
higher-order coactivities, a readout neuron in a mereological schema $\mathcal{M}$ learns to detect ever larger groups of cells that together 
inhibit its activity (Figure~\ref{Figure7}D).

Similarly to the overlap orders in $\mathcal{T}$, the cover relationships, as a rule, cannot be deduced from a single coactivity event. Thus, 
these relationships represent integral information that can be viewed as the $\mathcal{M}$-integrals which characterize the large-scale 
topology of a space. We are currently unaware of additional mereological algorithms that would allow large-scale computations of the 
environment's global topological characteristics, similar to computing the homological invariants in a simplicial schema. Nevertheless, a mereological 
schema encodes an important type of topological information, which may be used in physiological neural networks to represent spatial maps.

In general, covering relationships can be established between arbitrary (including multiply connected) regions. As a result, the number of 
possible combinations of covered and covering regions dramatically increases. Even if the covered region $V =\{d_1, d_2, ... , d_l\}$ is 
spatially ``bundled'' (e.g., if each pair $d_i$ and $d_j$ is coactive at some moment of time, so that $V$ forms a connectivity clique) the 
selection of possible covering regions remains very large. Therefore, in order to test the development of cover relationships in time, we 
opted to limit our study to neuron pairs covering an individual neuron ($k =2$ and $l =1$).
\begin{figure}
{\includegraphics[scale=0.9]{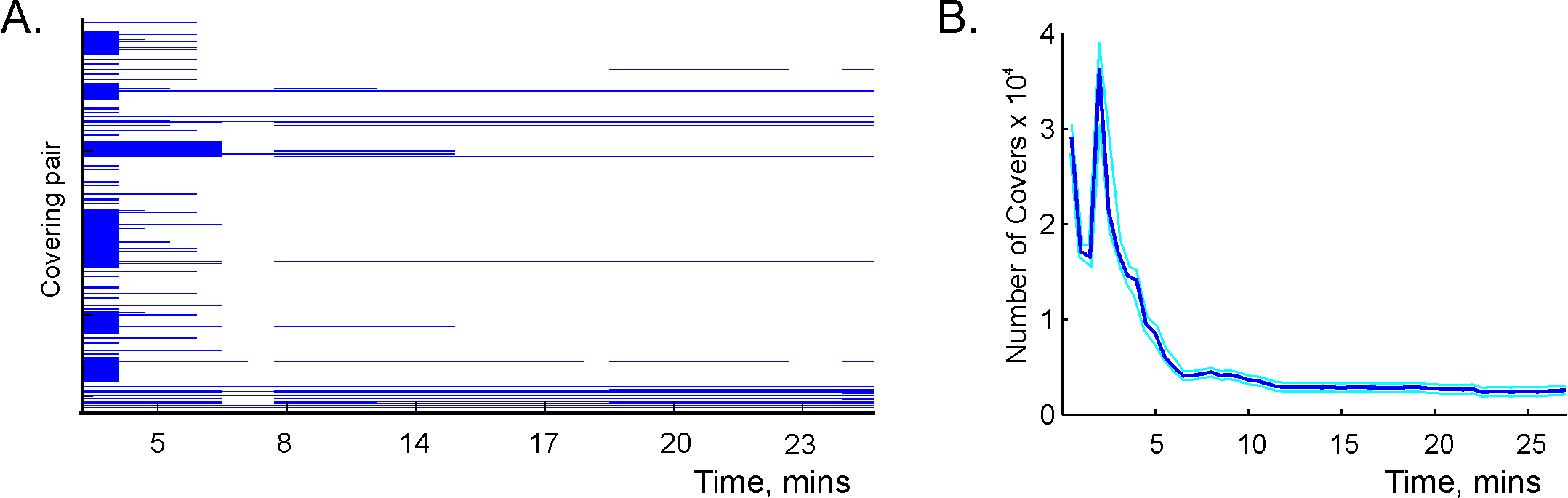}}
{\caption{\textbf{Learning the mereological schema}. \textbf{A}. Time development of the covering relationships: a pair of covering place 
cells, $U = \{c_i, c_j\}$, and a covered cell, $V =\{d_k\}$. Each line corresponds to a specific choice of $U$ and $V$. Each line begins as soon 
as the covering relationship is detected and stops as soon as it is violated, that is, as soon as the readout neuron shown in Figure~\ref{Figure7}D 
would fire. Note that the majority of relationships are short-lived: a large number of spurious relationships are detected at the beginning of 
exploration. After about seven minutes the majority of them disappear, similar to the behavior of topological loops computed in the simplicial 
schema shown in Figure~\ref{Figure6}B. This diagram shows about $1 \% $ of the detected pairs, selected at random. 
\textbf{B}. The number of detected cover relationships between pairs of place cells and a single place cell as a function of time.}\label{Figure8}}
\end{figure}

The results of simulations show that the time required to learn second-order covering relationships in $\mathcal{M}$ is comparable to the 
learning times in the graph schema $\mathcal{G}$ (Figure~\ref{Figure8}). As spatial exploration begins, a large number of transient covering 
relationships is produced because of insufficient spiking data. With accumulating spike trains most cover relationships become violated, so that 
the number of surviving relationships quickly drops. As the animal completes its first turn around a central hole of the environment 
(Figure~\ref{Figure1}B), a new set of (mostly transient) relationships is injected into the schema which produces the peak shown in 
Figure~\ref{Figure8}B. Subsequently, the number of cover relations steadily diminishes to about 200 pairs, which corresponds to a saturated 
schema. This result reflects qualitative behavior of higher order covering relationships, though a full implementation of the algorithm for the 
higher-order covering combinations ($k$, $l > 1$) is computationally substantially more complex.

\subsection{Complex Relations and the \textsf{RCC} Schema $\mathcal{R}$}
\begin{figure}
{\includegraphics[scale=0.9]{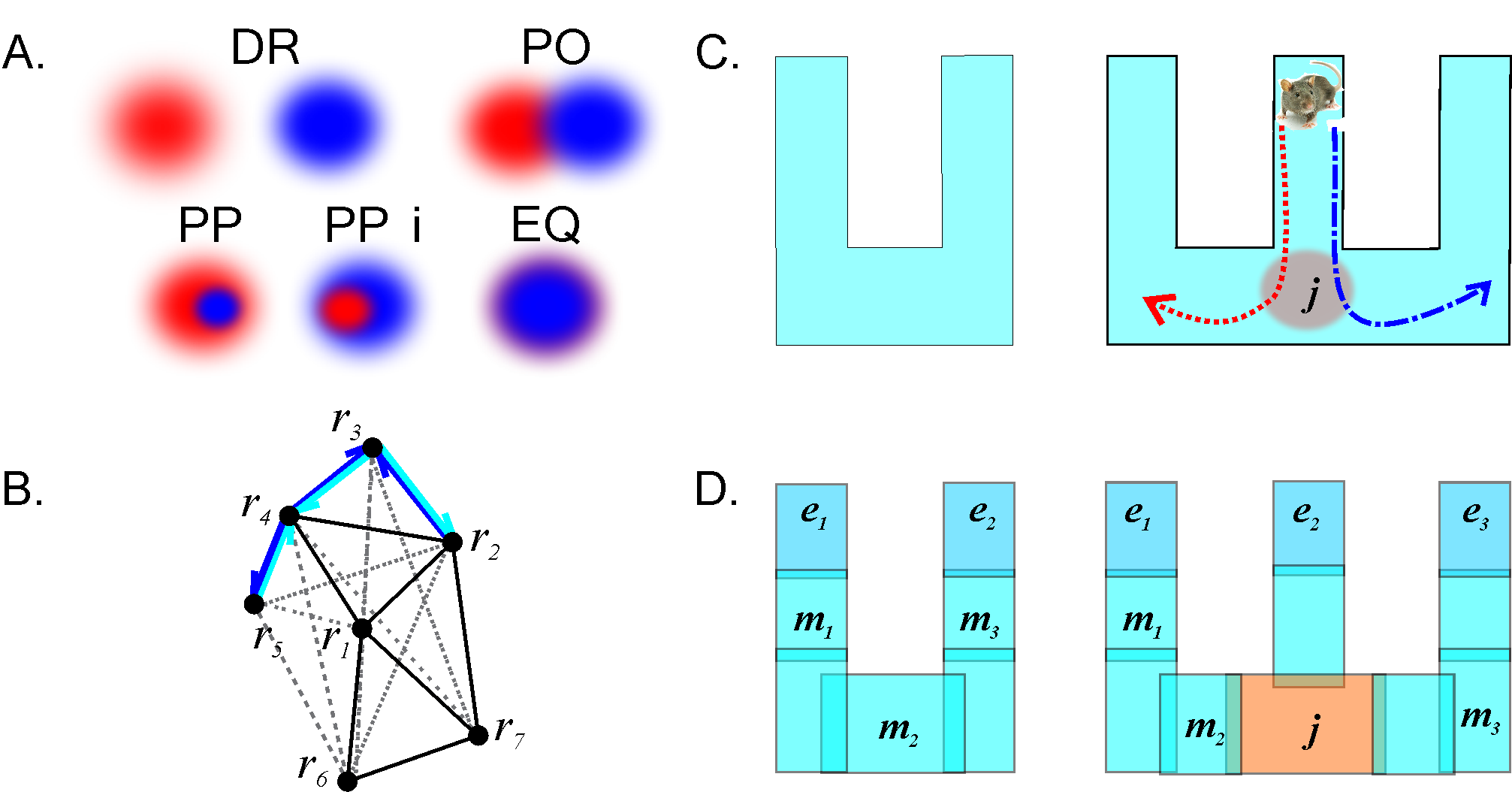}}
{\caption{\textbf{Illustration of \textsf{R\textbf{C}\textbf{C}5} schema}. \textbf{A}. \textsf{RCC5} relationships: five logically possible pairwise 
relations: ``$x$ is discrete from $y$'' (denoted as \textsf{DR}), ``$x$ partially overlaps with $y$'' (\textsf{PO}), ``$x$ is a proper part of $y$'' 
(\textsf{PP}), ``$y$ is a proper part of $x$'' (\textsf{PPi}), and ``$x$ is identical to $y$'' (\textsf{EQ}). \textbf{B}. An \textsf{RCC5} schema,
$\mathcal{R}_5$, of the spatial map from Figure~\ref{Figure2}B. For convenience, the \textsf{DR} connections are shown with gray dashed 
lines. The structure of the rest of the relations produces a graph similar to the one shown in Figure~\ref{Figure2}\textbf{C}. The black lines 
indicate \textsf{PO} connections. Cyan and blue arrows show the \textsf{PP} and \textsf{PPi} connections, respectively. \textbf{C}. A $U$-track 
having two dead ends and a $W$-track having three dead ends and a junction, $j$, marked in red. Every time the rat visits the junction point it 
must choose between the left and the right turn, indicated by the red and blue trajectories, respectively. \textbf{D}. Topological relationships 
between regions on a $U$- and a $W$-track that allow capturing the tracks’ qualitative geometries. The endpoints, $e_1$, $e_2$ and $e_3$ 
are regions that overlap with only one other region. The midpoints, $m_1$, $m_2$ and $m_3$, overlap with two regions and the junction 
overlaps with three regions.}\label{Figure9}}
\end{figure}

Qualitative Space Representations (\textsf{QSR}) are discrete, region-based versions of the conventional point-set theoretical geometries and 
topologies \cite{Hazarika} used to formalize ``intuitive'' qualitative spatial reasoning \cite{Gotts1,Renz,Cohn}, and thus are particularly important 
for modeling cognitive representations of space \cite{Knauff,Goodrich,Wallgrun,Zeithamova}. Important examples of \textsf{QSR}s are the Region 
Connection Calculi (\textsf{RCC})---formal logical theories based on a family of binary topological relations between regions \cite{Cui}. For example, 
the most basic \textsf{RCC} theory, \textsf{RCC5}, which applies to the case of regions with fuzzy boundaries, is built using the five relations shown 
in Figure~\ref{Figure9}A: disconnect (\textsf{DR}), partial overlap (\textsf{PO}), proper part and its inverse (\textsf{PP} and \textsf{PPi}), and equality 
(\textsf{EQ}) \cite{Cohn}. In terms of these relations, the arrangement of regions shown on Figure~\ref{Figure2}B is described by the following set 
of \textsf{RCC5} relationships: 
$P =\{\textsf{PO}_{12}, \textsf{PO}_{14}, \textsf{PO}_{16}, \textsf{PO}_{17}, \textsf{PPi}_{23}, \textsf{PO}_{24}, \textsf{PO}_{27}, \textsf{PP}_{32}, 
\textsf{PP}_{34}, \textsf{PPi}_{43}, \textsf{PPi}_{45}, \textsf{PO}_{46}, \textsf{PP}_{54}, \textsf{PO}_{67}; \textsf{DR}$ for all other pairs$\}$ 
(Figure~\ref{Figure9}B).  
More elaborate \textsf{RCC} calculi can capture tangencies \cite{Cui}, convexity \cite{Cohn}, qualitative directions \cite{Li}, and distances \cite{Gerevini} 
as well as complex hierarchies of all these relationships \cite{Lehmann}. As a result, \textsf{RCC} methods can capture not only standard topological 
signatures of spaces, such as loops and holes \cite{Gotts2}, but also more subtle qualitative features, such as branching points, linear sections, and dead 
ends. These qualitative features produce fundamental differences in spatial reasoning required for navigating the corresponding environments. 
For example, the junction point on the W-tracks, which are often used in behavioral experiments (Figure~\ref{Figure9}C), forces an animal to choose 
between a right or left turn, which is reflected in the place cell code \cite{Frank2,Huang}. The \textsf{RCC5} theory allows capturing such features, e.g., 
distinguishing between the $U$- and $W$-tracks, which, from the perspective of algebraic topology, are but contractible manifolds (Figure~\ref{Figure9}D).
\begin{figure}
{\includegraphics[scale=0.87]{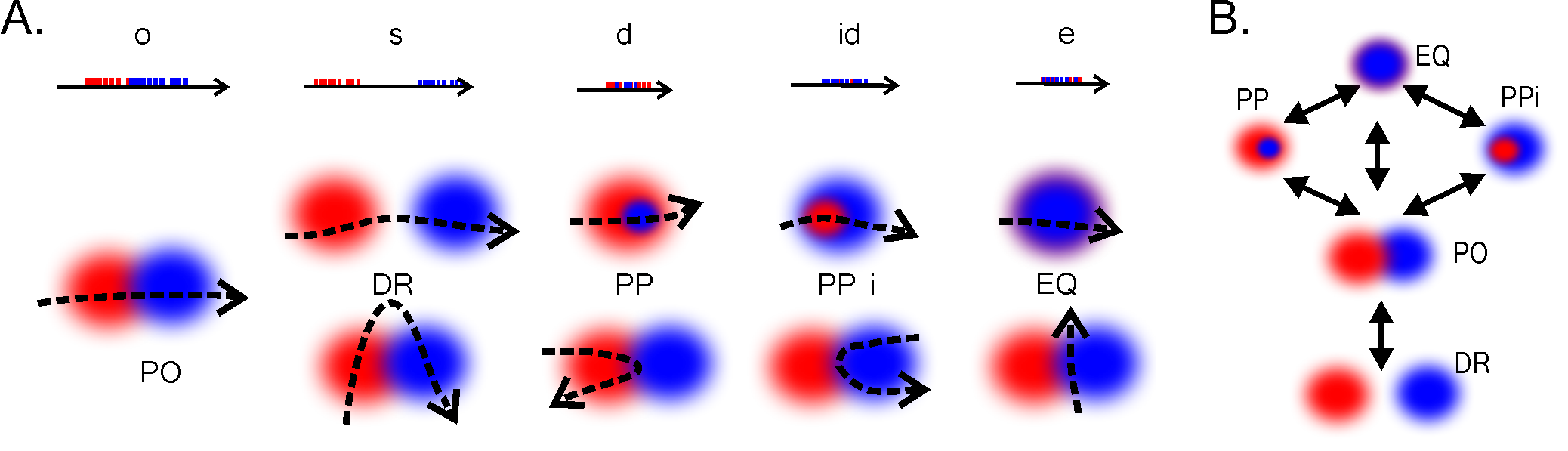}}
{\caption{\textbf{Temporal vs. spatial relationships}. \textbf{\textbf{A}}. Temporal relationships between the spike trains (\textsf{o} – overlap, 
\textsf{s} – separation, \textsf{d} – during, \textsf{id} – inverse \textsf{d} and \textsf{e} – equal \cite{Ligozat}) and the corresponding spatial relationships. 
The relationships \textsf{DR}$_{xy}$, PP$_{xy}$, \textsf{PPi}$_{xy}$ and \textsf{EQ}$_{xy}$ between the regions can be imitated by partial overlap, 
depending on the shape of the trajectory, which shows that these relations cannot be directly deduced from the spike train structure. \textbf{B}. The 
transitions between the \textsf{RCC5} relationships, showing the immediate conceptual neighborhood (continuity table) structure of \textsf{RCC5}. 
These are the possible sequences of gradual transformation of the \textsf{RCC5} relationships. For example, if at some moment of time two regions, 
$x$ and $y$, were disconnected (\textsf{DR}$_{xy}$) then this relationship cannot instantly jump to a containment relationship (PP$_{xy}$ or 
\textsf{PPi}$_{xy}$) without going through, at least instantaneously, the partially overlapping (\textsf{PO}$_{xy}$) relationship.}\label{Figure10}}
\end{figure}

To model spatial learning based on a specific \textsf{RCC} approach, one can construct an \textsf{RCC} schema, in which the readout neurons are trained 
to recognize the appropriate set of binary relationships. However, an important aspect of \textsf{RCC} constructions is that the set of relationships that can 
be simultaneously imposed on a set of regions is restricted \cite{Bennett,Renz}.
For example, if $x$ and $y$ partially overlap and $y$ is a proper subset of $z$, then $z$ and $x$ must have a non-null intersection and $z$ cannot be a 
subset of $x$. Therefore, we define an \textsf{RCC} schema $\mathcal{R}$ as a schema with a set of \textit{consistent} \textsf{RCC} relations between regions.

To model the process of physiological learning in the \textsf{RCC5} ($\mathcal{R}_5$) schema, we trained five types of readout neurons to recognize the 
five \textsf{RCC5} relationships, starting from the initial \textsf{DR} relationship. This however requires more complex algorithms than in $\mathcal{G}$ 
and $\mathcal{T}$ schemas: while the partial temporal overlap can always be interpreted as partial spatial overlap, other temporal relationships cannot 
be uniquely assigned to a spatial \textsf{RCC5} relation (Figure~\ref{Figure10}A). For example, passing through two partially overlapping regions along 
a particular trajectory can generate a temporal disconnect, a temporal cover, or a temporal equality relationship between two spike trains which can be 
mistaken for spike trains produced by a \textsf{DR}, \textsf{PP}/\textsf{PPi}, or \textsf{EQ} relationship, respectively. Because of this ambiguity, the spiking 
activity of the presynaptic cells in the cell assemblies produced during individual runs through a pair of place fields can invoke different interpretations of the 
spatial relationships. Thus, learning a $\mathcal{R}_5$ schema rests on encoding, at each moment, the best guesses for the relationships between pairs of 
regions and then updating them based on the available spiking history and the qualitative analogue of continuity constraints, as shown in Figure~\ref{Figure10}B. 

\begin{figure}
{\includegraphics[scale=0.9]{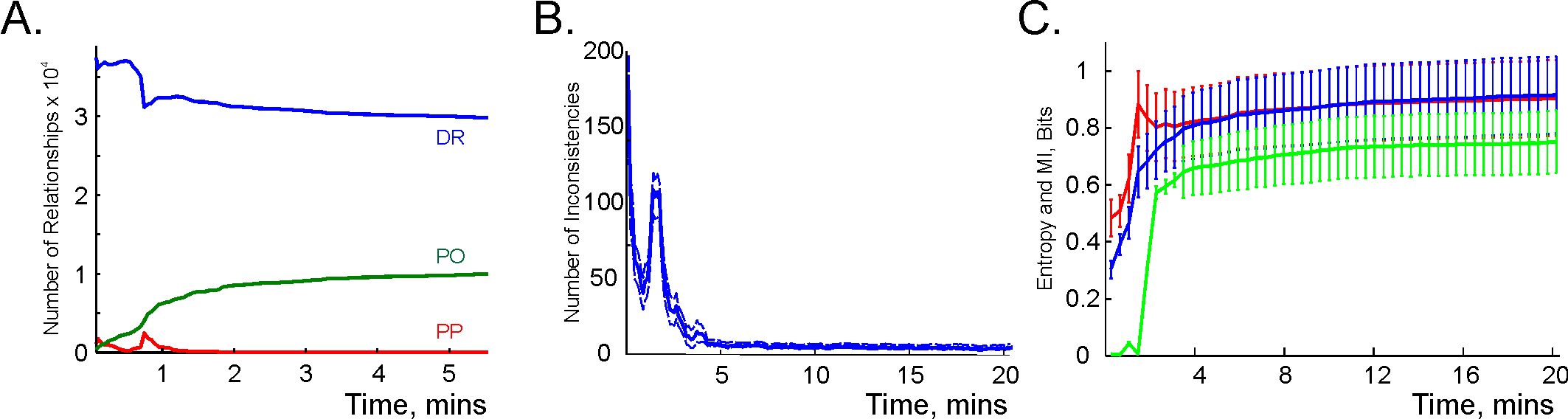}}
{\caption{\textbf{Learning the \textsf{RCC5} schema}. \textbf{A}. Evolution of \textsf{RCC5} relationships: at the beginning of the exploration 
of the new space, the regions are mostly disconnected and the partial overlap of relationships is accumulated over time.  Similarly to the graph 
schema, the number of all types of relationships saturates in about five minutes. \textbf{B}. The number of inconsistencies in the randomly 
initialized relational network is higher at the beginning of exploration and decays to a low, steady level by the time the number of relationships 
stabilize. \textbf{C}. The entropy of the relations encoded in the schema (red), the entropy of the \textsf{RCC5} relationships in the map (blue), 
and the mutual information (green) between them saturate at a similar time scale.}\label{Figure11}}
\end{figure}

In our simulations, the relationships evolved rapidly and saturated within about $T_{\mathcal{N}}  \approx 4$ minutes from the onset of the exploration 
(Figure~\ref{Figure11}A). Figure~\ref{Figure11}B shows that at the beginning of the exploration, the number of inconsistencies between independently 
trained readout neurons is high. Subsequently, their number quickly diminishes as the information about coactivity is acquired. An increase in the number 
of \textsf{PP} relationships in Figure~\ref{Figure11}A produces a splash of inconsistencies occurring at about 3 minutes, which is at the time when the 
animal completes its first turn around the central hole. This phenomenon has the same origin as the splash of transient cover relationships occurring in 
the mereological schema $\mathcal{M}$ (Figure~\ref{Figure8}). 

As the statistical information about place cell coactivity accumulates, a stable set of \textsf{RCC5} relationship emerges. The schema's specific 
entropy, defined using the probabilities of observing all five relationships, saturates about the same time, $T_{\mathcal{N}}  \approx 4$ minutes. 
The entropy of the \textsf{RCC5} relationships between the representing regions in the map $M_X (\mathcal{R})$ remains similar to the schema 
entropy during the entire course of learning, reaching the asymptotic value of $H \approx 0.84$ (Figure~\ref{Figure11}C). Moreover, the mutual 
information between the map and the schema increases with the acquisition of information in a way similar to the case of graph schema 
$\mathcal{G}$ but unlike the case of simplicial schema $\mathcal{T}$ (cf. Figures~\ref{Figure11}C, \ref{Figure3}C and \ref{Figure5}E). Once 
again, this data indicates that spatial maps built on regions with diffuse boundaries may better reflect the nature of the encoded regions.
In the meantime, the integrals of the $\mathcal{R}_5$ schema, i.e., the combinations of \textsf{RCC5} relationships that represent the junction 
and the endpoints on the $W$ track, emerge from neuronal spiking in under $T_{\min}  \approx 2$ minutes---much sooner than the readout 
neurons in $\mathcal{R}_5$ network are trained.

\section{DISCUSSION}
\label{section:discussion}

We have presented a framework for integrating the place cells' spiking information into a global map of space, implemented via simple cell 
assembly neural networks, wired to encode spatial relationships. The approach is motivated by the experimental results \cite{Alvernhe1,Alvernhe2,Poucet1,Shapiro,Wu,Chen1,eLife,Lever,Leutgeb,Wills,Touretzky,Gothard1,Gothard2} and by the theoretical 
models proposed in \cite{Dabaghian,Arai} and in \cite{cGraph,Trullier,Chrastil}. From the perspective of the current approach these models 
are particular implementations of the simplicial and the graph schema, respectively; the mereological and the \textsf{RCC} schemas are new---to 
our knowledge, such models have not yet been considered. 

In the following we outline several important aspects of this framework and provide a general context for the model.

\textbf{4.1 Emergence of the memory map in schemas}. There is a clear parallel between a coherent representation of space emerging from the
integrated inputs of many individual neurons and a continuous state of matter (e.g., a solid or a liquid) emerging from the collective dynamics of 
molecules. From a descriptive point of view, the common element in both phenomena is that neither macro-system can be reduced to a trivial 
aggregation of the properties of its elementary constituents. Even when the properties at both the microscopic and macroscopic levels are well 
understood, it can be difficult to correlate the properties at one level with those at the other. 
For example, the measurement of the macroscopic properties of a liquid does not allow one to determine its molecular structure. Conversely, a 
detailed description of the properties of a water molecule does not explain directly key phenomena of the physics of water. In physics, the solution 
to this problem historically proceeded from a simplified, phenomenological models, which bridged the gap between the microscopic and macroscopic 
levels of matter. In a similar way, the present discussion offers a testbed model with which to bridge the gap between place cells and the large-scale 
spatial map.

\textbf{4.2 Topological spatial maps}. Topological maps have several biological advantages over geometric (or topographic \cite{Chen2}) maps, which 
follow from the qualitative nature of topological relationships \cite{Chen1}. First, natural environments are dynamic, so that it is often impossible for an 
animal to know when and how its navigational task may change. Hence acquiring a qualitative map based on the invariants of the space of an environment, 
may be biologically more effective than spending time on producing a computationally costly precise answer from mutable relationships between 
dynamic cues.

One implication of this hypothesis is that in morphing environments the place fields will retain the pattern of topological connectivity and may adjust 
their shapes in order to compensate for the deformation of the representing space. This hypothesis is supported by experiments which 
demonstrate that place fields maintain their relative configurations in morphing environments 
\cite{Muller2,Colgin,Lever,Leutgeb,Wills,Touretzky,Gothard1,Gothard2} and that place cell coactivity pattern in an animal traversing remains invariant 
over a significant range of geometric transformations \cite{eLife, Diba}. If the map is Cartesian, i.e., based on precise coordinates, distances, angles 
and so forth, such deformation can be achieved by redrawing the place fields at each stage of the deformation, via some complex path integration 
mechanism \cite{McNaughton1,Poucet1,Alvernhe1,Goodrich,Etienne,McNaughton2,Poucet2}. From the topological perspective, the observed deformation
of the place fields is simply a result of projecting the same stable neuronal map into a morphing environment, which does not require extra computations
and hence may be biologically more plausible.

\textbf{4.3. Schemas constrain the generation of intrinsic sequences in the hippocampus}. Place cells become active in temporal sequences that either 
match with or are inverse to the spatial ordering of their place fields during the active, resting, or sleep states. Initially, temporal sequences were 
observed after or during the recording of the place fields, leading various authors to suggest that the observed temporal sequences were a replay of
sequences imprinted by sensory inputs \cite{Foster,Louie}. More recent experiments have observed temporal sequences that corresponded to trajectories
along which the animal had never traveled \cite{Gupta}. Furthermore, experiments have revealed that temporal sequences observed before the animal 
entered an environment for the first time were predictive of the place field sequence measured later \cite{Dragoi1,Dragoi2}. These observations strongly 
suggest that temporal sequences are not merely replays of previously imprinted sequences \cite{Gupta,Cheng1}. The better interpretation is that sequences 
are drawn from a pool of sequences that are intrinsic to the hippocampal network and this network structure gives rise to the location of place fields 
\cite{Cheng2,Buhry,Azizi}. The CRISP (for Context Representation, Intrinsic Sequences, and Pattern completion) theory goes further to argue that the 
intrinsic sequences in the hippocampus are crucial for the storage of episodic memories \cite{Cheng2}. However, this theory does not explain the origin 
or properties of such sequences.

In the schema framework, all neural activity produced in the hippocampus has to be consistent with its schema. For example, in the graph schema,
spontaneously replayed sequences of neural activity would have to be consistent with the connectivity of the graph. In other words, a cell $c_i$ may 
fire a spike after cell $c_j$ only if the relationship $\rho_{ij} = (r_i, r_j)$, or a chain of intermediate relationships $\rho_{i i_1}$, $\rho_{i_1 i_2}$, ..., 
$\rho_{i_{n-1}i_{n}=j}$, is present in $P_R$. 
Other schemas impose different constraints on which sequences can be produced, and the elements in the sequences may be ensembles of place cells, 
rather than single cells. In other words, schemas serve as ``topological templates'' off which sequences are generated. 

Physiologically, this implies that the hippocampal network that implements a particular schema can produce sequences with specific ``grammar'' 
which may not have been directly imprinted or previously produced by the network. In fact, such offline state sequences of place cell activations, 
which the animal had never experienced, were recently observed in the experiments \cite{Dragoi1,Dragoi2}. Moreover, these sequences were 
consistent with the topology of the spatial environment \cite{Wu}. Thus, schemas can explain the intrinsic sequences postulated by CRISP theory 
as well as in preplay and replay. This intimate relationship between spontaneous sequences and schemas may be exploited in future investigations 
in order to infer the schema based on recordings of sequences or  to predict the properties of intrinsic sequences from a given schema.

\textbf{4.4. Spatial vs. non-spatial memories.} 
The hippocampus has been suggested to encode both spatial and nonspatial memories \cite{Eichenbaum2,Konkel,Shrager,Soei,Eichenbaum3}. 
For example, it plays a key role in the ability to remember visual, odor, action and memory sequences, and to put a specific memory episode 
into the context of preceding and succeeding events, as well as the ability to produce complete memory sequence from a single structured input \cite{Wood,Fortin1,Fortin2,Sauvage}. The topological view on the hippocampal spatial representations \cite{Dabaghian,eLife,Dabaghian1} provides 
a common framework for understanding both spatial and nonspatial memory functions as manifestations of a single mechanism, which simply 
produces a topological arrangement of memory elements, irrespective of the nature of their content. According to this view, there is no principal 
difference between the internalized topological map of spatial locations and a topological map of memory sequences in the mnemonic domain. 

\textbf{4.5. Connections to experiment.} 
Given the place cells' spiking parameters and a hypothesis about how the downstream neurons might process place cell (co)activity, a schematic 
computation can be used to assess the effectiveness of the corresponding spatial learning mechanism: how much time will be required to map a 
space, how many integrals can such mapping produce, how quickly these integrals will emerge and how stable they will be. This scope of computations 
suggests a possibility for experimental verifications of the proposed framework. For example, a decline in spatial learning caused by neurodegenerative 
diseases (e.g., in Alzheimer's rat models), by aging or by psychoactive substances is assessed in behavioral experiments in terms of the extra times
required to learn various memory tasks. On the other hand, such cognitive changes are associated with changes in the place cell spiking parameters
\cite{Cacucci,Robbe2,Silvers,Gerrard,Robitsek,WilsonX}. It may therefore be possible to compare the downturn of spatial memory observed in 
topological learning tasks \cite{Alvernhe1,Poucet1} with the increase of the learning time(s) estimated via a particular schema model for the same 
change in spiking variables.
Another alternative was suggested to us by our recent studies of hippocampal mapping of 3D spaces in bats, using two types of simplicial schemas. 
The results suggest that in the 3D case, the readout neurons in the place cell assemblies should operate by integrating synaptic inputs over working 
memory periods, rather than detecting coactivities on synaptic plasticity timescale \cite{Kentaro}. Of course, until these predictions are proved or 
disproved experimentally, their value is discussable; meanwhile, the schema approach allows theoretical reasoning and generates predictions about hippocampal neurophysiology.

\section{Methods}
\label{section:methods}

\textbf{Place cells}. Spiking is produced by the rat's movement through the environment covered by the place fields (Figure~\ref{Figure1}A-B). 
The Poisson rate of the firing of place cells is a function of the animal's position $r(t)$ at time $t$,
$$\lambda_i(r)=f_ie^{-\frac{(r-r_i)^2}{2s_{i}^2}},$$
where $f_i$ is the maximal firing rate of cell $c_i$, $s_i$ defines the width of its firing field centered at $r_i$ \cite{Barbieri}. In an $N$-cell ensemble, 
the parameters $f_i$, and $s_i$, $i =1, ..., N$ are modeled as random variables drawn from stationary unimodal distributions characterized by their 
respective means ($f$ and $s$) and standard deviations (see Figure~\ref{Figure1} and Methods in \cite{Dabaghian}). For the computations we used
an ensemble with $N=200$ neurons, with mean firing rate $f=12$ Hz and the mean place field size $s=20$ cm. Larger ensembles typically affect the numerical values of the computed quantities, but not the essence of the phenomena described in the paper. This spiking is modulated by theta-oscillations, which are a subcortical EEG cycle in the hippocampus with a 
frequency of $\sim 8$ Hz (for details see \cite{Arai}). 

\textbf{Learning Algorithm}. 
The physiological processes responsible for emergence of cell assemblies with readout neurons trained to integrate presynaptic inputs and to produce 
a particular response that ``actualizes'' the information encoded by the place cell coactivity are complex and multifaceted \cite{Buzsaki}. For example, 
the readout neurons that encode place field overlap must identify a group of place cells and learn to respond to the coactivity of this specific group. 
However, what matters for our study, are the qualitative results of this process: the number of readout neurons , the order of the coactivity detected 
by these neurons, how this order grows in a typical cell assembly during the learning process and so forth. 
Therefore, we set aside a neural network simulation of schema learning and employ the following schematic, phenomenological algorithm:

If a relationship $\rho$ of an appropriate type is detected, then:
\begin{enumerate}
\item if $\rho$ is already listed in $P_R$, ignore;
\item else if $\rho$ can be inferred from the known relationships, ignore;
\item if $\rho$ provides nontrivial information, then add $\rho$ to $P_R$.
\item if the known relationships can be inferred from $\rho$, then remove the redundant relationships.
\item continue
\end{enumerate}

Steps 2 and 3 ensure that only the highest order relationships are kept in the schema, eliminating redundant, lower-order relationships. At the beginning, 
every state of the readout neurons can be empty and trained as the simulated animal explores a novel environment, or these states can be randomly 
initialized and then relearned. The transitions between the readout neuron types may be regarded as a rudimentary, schematic model of the synaptic 
plasticity mechanisms. In novel environments, place fields stabilize in about four minutes \cite{Brown2}, even though cognitive learning of the environment 
may take days or even weeks \cite{Frank3}. This implies that the readout neurons can be trained using constant spiking parameters $f_i$ and $s_i$.

\textbf{Temporal relationships} between the spike trains and the physiological mechanisms underlying the downstream neurons' readout process are in
general very complex. For the sake of simplicity, we consider only the rate-based representation of neuronal activity \cite{Ahmed}, which allows for a 
variety of possibilities for encoding relationships. Such relationships may entail that the firing rates of the pre- and postsynaptic neurons may be required 
to fall within a particular interval of values and the period of activity of a neuron $c_i$ may be required to precede, to follow, or to overlap with the activity 
of a neuron $c_j$ by a certain minimal, maximal or fixed amount of time \cite{Ligozat}. The present analysis works from the three mutually exclusive logical 
possibilities for the activity of any two neurons $c_i$ and $c_j$ : 

\begin{enumerate}
\item there is an empty intersection of activity, i.e., the two cells are active at different times;
\item there is a non-null intersection of activity, i.e., their activities overlap;
\item the activity of one cell is a proper subset of the other cell, i.e., the activity of one cell occurs entirely within the timespan of the activity of the other cell.
\end{enumerate}
The time window for defining the co-activity of two or more cells is two $\theta$-periods \cite{Arai,Mizuseki}.

Schema entropy and mutual information. For each relationship $\rho_k$ of the schema we computed its normalized frequency of appearance $p_k$ and 
evaluated the resulting specific entropy,
$$H=-\Sigma_k p_k \log_2 p_k .$$
The specific entropy for the corresponding spatial map was evaluated by identifying the relationships $\rho_{k'}$ that obtain between the corresponding 
representing regions and computing their appearance probabilities $p_{k'}$. Following the trajectory of the animal (Figure~\ref{Figure1}B), we could 
also detect the joint probability of appearance $p_{k,k'}$ of a given pair of relationships, both in the schema as well as in the map $(\rho_k,\rho_{k'})$, 
and then compute their mutual information between the map and the schema,
$$MI=-\Sigma_k\Sigma_{k'}p_{k,k'}\log_2\frac{p_{k,k'}}{p_k p_{k'}}.$$

The computational software used for topological analysis is JPlex, an open-source package implementing Persistent Homology Theory methods developed 
by the Computational Topology group at Stanford University \cite{JPlex}.

\section{Acknowledgments}
\label{section:acknow}
We thank Robert Phenix, Vicky Brandt and Loren Frank for helpful comments.
The work was supported in part by Houston Bioinformatics Endowment Fund (A.B. and Y.D.), the W. M. Keck Foundation 
grant for pioneering research (A.B. and Y.D.) and by the NSF 1422438 grant (A.B. and Y.D.), and by the German Research 
Foundation (Deutsche Forschungsgemeinschaft, DFG): SFB 874, project B2 (S.C.), a grant from the Stiftung Mercator (S.C.).

\end{document}